\documentclass[fleqn,twoside]{article}
\usepackage{espcrc2}

\usepackage{graphicx}

\newcommand{\AmS}{{\protect\the\textfont2
  A\kern-.1667em\lower.5ex\hbox{M}\kern-.125emS}}

\title{Two mixing angle description of radiative $VP\gamma$ decays%
         \thanks{UAB--FT--589 report.
                To appear in the proceedings of the QCD 05:
	       12th High-Energy Physics International Conference
	       in Quantum Chromodynamics,
	       4-8th July 2005 Montpellier (France).}}

\author{R. Escribano\address{Grup de F\'{\i}sica Te\`orica and IFAE, 
        Universitat Aut\`onoma de Barcelona,\\ 
        E-08193 Bellaterra (Barcelona), Spain}%
	\thanks{Work partly supported by the Ramon y Cajal program,
                        the Ministerio de Ciencia y Tecnolog\'{\i}a and FEDER, FPA2002-00748EU,
                        and the EU, HPRN-CT-2002-00311, EURIDICE network.}}
       
\begin{document}

\begin{abstract}
\noindent
An analysis of various radiative $VP\gamma$ decays is performed using
the two mixing angle description of the $\eta$-$\eta^\prime$ system.
The agreement is excellent.
Our results are expressed both in the ``octet-singlet''  and in the ``quark-flavour'' basis.
It turns out that at the present experimental accuracy, 
the two angles are significantly different in the former, but not in the latter basis.
\end{abstract}

\maketitle

\section{INTRODUCTION}
\noindent
The mixing pattern of the pseudoscalar decay constants associated to the $\eta$-$\eta^\prime$ system
is usually described in terms of a single mixing angle \cite{Ball:1995zv}.
However, this is not the most general mixing scheme since two axial currents,
the eight component of the octet and the singlet,
can couple to the two physical states, the $\eta$ and $\eta^\prime$.
Therefore, a mixing scheme consisting of two mixing angles should be used instead of the simplest one mixing angle description.
The reason for not using the two mixing angle scheme from the beginning has been the absence of a well established framework able to give some insight on the values of the two different angles and their related decay constants.
Now with the advent of Large $N_c$ Chiral Perturbation Theory ($\chi$PT) \cite{Kaiser:2000gs},
where the effects of the pseudoscalar singlet are treated in a perturbative way,
the new two mixing angle scheme receives theoretical support.
The aim of this work is to perform an updated phenomenological analysis
of various radiative $VP\gamma$ decays using the two mixing angle description of the
$\eta$-$\eta^\prime$ system.
The analysis will serve us to check the validity of the two mixing angle scheme and its improvement over the standard one angle picture.
Our analysis also tests the sensitivity to the mixing angle schemes:
\emph{octet-singlet} versus \emph{quark flavour} basis.

\section{NOTATION}
\label{notation}
\noindent
The decay constants of the $\eta$-$\eta^\prime$ system in the octet-singlet basis
$f_P^a\ (a=8,0; P=\eta,\eta^\prime)$ are defined as
\begin{equation}
\label{decaycon}
\langle 0|A_\mu^a|P(p)\rangle=i f_P^a p_\mu\ ,
\end{equation}
where $A_\mu^{8,0}$ are the octet and singlet axial-vector currents whose
divergences are
\[
\label{divaxialos}
\begin{array}{l}
\partial^\mu A_\mu^8=\frac{2}{\sqrt{6}}
(m_u\bar u i\gamma_5 u+m_d\bar d i\gamma_5 d-
2m_s\bar s i\gamma_5 s)\ ,\\[2ex]
\partial^\mu A_\mu^0=\frac{2}{\sqrt{3}}
(m_u\bar u i\gamma_5 u+m_d\bar d i\gamma_5 d+m_s\bar s i\gamma_5 s)+\\[2ex]
\qquad\qquad
+\frac{1}{\sqrt{3}}\frac{3\alpha_s}{4\pi} G_{\mu\nu}^a\tilde G^{a,\mu\nu}\ ,
\end{array}
\]
where $G^a_{\mu\nu}$ is the gluonic field-strength tensor and
$\tilde G^{a,\mu\nu}\equiv\frac{1}{2}\epsilon^{\mu\nu\alpha\beta}G^a_{\alpha\beta}$
its dual.
The divergence of the matrix elements (\ref{decaycon}) are then written as
\begin{equation}
\label{divdecaycon}
\langle 0|\partial^\mu A_\mu^a|P\rangle=f_P^a m_P^2\ ,
\end{equation}
where $m_P$ is the mass of the pseudoscalar meson.

\noindent
Each of the two mesons $P=\eta, \eta^\prime$ has both octet and singlet components,
$a=8, 0$.
Consequently, Eq.~(\ref{decaycon}) defines \emph{four independent} decay constants.
Following the convention of Refs.~\cite{Leutwyler:1997yr,Kaiser:1998ds}
the decay constants are parameterized in terms of two basic decay constants
$f_{8}, f_{0}$ and two angles $\theta_{8}, \theta_{0}$
\begin{equation}
\label{defdecaybasisos}
\left(
\begin{array}{cc}
f^8_\eta & f^0_\eta \\[1ex]
f^8_{\eta^\prime} & f^0_{\eta^\prime}
\end{array}
\right)
=
\left(
\begin{array}{cc}
f_8 \cos\theta_8 & -f_0 \sin\theta_0 \\[1ex]
f_8 \sin\theta_8 &  f_0 \cos\theta_0
\end{array}
\right)\ .
\end{equation}

\noindent
Analogously, in the quark-flavour basis the decay constants are parameterized in terms of
$f_{q}, f_{s}$ and $\phi_{q}, \phi_{s}$ \cite{Feldmann:1999uf}:
\begin{equation}
\label{defdecaybasisqf}
\left(
\begin{array}{cc}
f^q_\eta & f^s_\eta \\[1ex]
f^q_{\eta^\prime} & f^s_{\eta^\prime}
\end{array}
\right)
=
\left(
\begin{array}{cc}
f_q \cos\phi_{q} & -f_s \sin\phi_{s}\\[1ex]
f_q \sin\phi_{q} &  f_s \cos\phi_{s}
\end{array}
\right)\ ,
\end{equation}
and the non-strange and strange axial-vector currents are defined as
\[
\label{divaxialqfos}
    \begin{array}{l}
    A_\mu^q=\frac{1}{\sqrt{2}}(\bar u\gamma_{\mu}\gamma_{5}u+\bar d\gamma_{\mu}\gamma_{5}d)
                    =\frac{1}{\sqrt{3}}(A_\mu^8+\sqrt{2}A_\mu^0)\ ,\\[2ex]
    A_\mu^s=\bar s\gamma_{\mu}\gamma_{5}s
                    =\frac{1}{\sqrt{3}}(A_\mu^0-\sqrt{2}A_\mu^8)\ .
    \end{array}
\]

\section{\boldmath{$VP\gamma$} FORM FACTORS}
\label{VPgamma}
\noindent
In order to predict the couplings of the radiative decays of lowest-lying vector mesons,
$V\rightarrow(\eta,\eta^\prime)\gamma$, 
and of the radiative decays $\eta^\prime\rightarrow V\gamma$, with $V=\rho, \omega, \phi$,
we follow closely the method presented in Ref.~\cite{Ball:1995zv} where the description of the
light vector meson decays is based on their relation with the $AVV$ triangle anomaly,
$A$ and $V$ being an axial-vector and a vector current respectively.
The approach both includes $SU_f(3)$ breaking effects and fixes the vertex couplings $g_{VP\gamma}$
as explained below.

\noindent
In that framework, one starts considering the correlation function
\begin{equation}
\label{corrfun}
\begin{array}{l}
i\int d^4x e^{iq_1 x}
\langle P(q_1+q_2)|TJ_\mu^{\rm EM}(x)J_\nu^V(0)|0\rangle=\\[2ex]
\qquad\qquad
=\epsilon_{\mu\nu\alpha\beta}q_1^\alpha q_2^\beta F_{VP\gamma}(q_1^2,q_2^2)\ ,
\end{array}
\end{equation}
where the currents are defined as
\begin{equation}
\label{defcurr}
\begin{array}{l}
J_\mu^{\rm EM}=\frac{2}{3}\bar u\gamma_\mu u-\frac{1}{3}\bar d\gamma_\mu d-
               \frac{1}{3}\bar s\gamma_\mu s\ ,\\[1ex]
J_\mu^{\rho,\omega}=
\frac{1}{\sqrt{2}}(\bar u\gamma_\mu u\mp\bar d\gamma_\mu d)\ ,
\quad J_\mu^\phi=-\bar s\gamma_\mu s\ .
\end{array}
\end{equation}
The form factors values $F_{VP\gamma}(0,0)$ are fixed by the $AVV$ triangle 
anomaly (one $V$ being an electromagnetic current), and are written in terms 
of the pseudoscalar decay constants and the $\phi$-$\omega$ mixing angle 
$\theta_V$ as
\begin{equation}
\label{FVPgamma00}
\begin{array}{l}
F_{\rho\eta\gamma}(0,0)=\frac{\sqrt{3}}{4\pi^2}
\frac{f^0_{\eta^\prime}-\sqrt{2}f^8_{\eta^\prime}}
{f^0_{\eta^\prime}f^8_\eta-f^8_{\eta^\prime}f^0_\eta}\ ,\\[1ex]
F_{\rho\eta^\prime\gamma}(0,0)=\frac{\sqrt{3}}{4\pi^2}
\frac{f^0_\eta-\sqrt{2}f^8_\eta}
{f^0_\eta f^8_{\eta^\prime}-f^8_\eta f^0_{\eta^\prime}}\ ,\\[1ex]
F_{\omega\eta\gamma}(0,0)=\frac{1}{2\sqrt{2}\pi^2}
\frac{(c\theta_V-s\theta_V/\sqrt{2})f^0_{\eta^\prime}-
      s\theta_V f^8_{\eta^\prime}}
{f^0_{\eta^\prime}f^8_\eta-f^8_{\eta^\prime}f^0_\eta}\ ,\\[1ex]
F_{\omega\eta^\prime\gamma}(0,0)=\frac{1}{2\sqrt{2}\pi^2}
\frac{(c\theta_V-s\theta_V/\sqrt{2})f^0_\eta-s\theta_V f^8_\eta}
{f^0_\eta f^8_{\eta^\prime}-f^8_\eta f^0_{\eta^\prime}}\ ,\\[1ex]
F_{\phi\eta\gamma}(0,0)=-\frac{1}{2\sqrt{2}\pi^2}
\frac{(s\theta_V+c\theta_V/\sqrt{2})f^0_{\eta^\prime}+
      c\theta_V f^8_{\eta^\prime}}
{f^0_{\eta^\prime}f^8_\eta-f^8_{\eta^\prime}f^0_\eta}\ ,\\[1ex]
F_{\phi\eta^\prime\gamma}(0,0)=-\frac{1}{2\sqrt{2}\pi^2}
\frac{(s\theta_V+c\theta_V/\sqrt{2})f^0_\eta+c\theta_V f^8_\eta}
{f^0_\eta f^8_{\eta^\prime}-f^8_\eta f^0_{\eta^\prime}}\ .\\[1ex]
\end{array}
\end{equation}
Using their analytic properties, we can express these form factors by
dispersion relations in the momentum of the vector current, which are then
saturated with the lowest-lying resonances:
\begin{equation}
\label{VMD}
F_{VP\gamma}(0,0)=\frac{f_V}{m_V}g_{VP\gamma}+\cdots\ ,
\end{equation}
where the dots stand for higher resonances and multiparticle contributions
to the correlation function.
In the following we assume vector meson dominance (VMD) and thus neglect these
contributions.

\noindent
The $f_V$ are the  leptonic decay constants of the vector mesons, and defined by
\begin{equation}
\label{lepdeccon}
\langle0|J_\mu^V|V(p,\lambda)\rangle=m_V f_V \varepsilon_\mu^{(\lambda)}(p)\ ,
\end{equation}
where $m_V$ and $\lambda$ are the mass and the helicity state of the vector meson.
The $f_V$ can be determined from the experimental decay rates \cite{Eidelman:2004wy} via
\begin{equation}
\label{GVee}
\Gamma(V\rightarrow e^+e^-)=\frac{4\pi}{3}\alpha^2\frac{f_V^2}{m_V}c_V^2\ ,
\end{equation}
with $c_V
=(\frac{1}{\sqrt{2}},\frac{s\theta_V}{\sqrt{6}},\frac{c\theta_V}{\sqrt{6}})$
for $V=\rho, \omega, \phi$.

\noindent
Finally, we introduce the vertex couplings $g_{VP\gamma}$, which are just the
on-shell $V$-$P$ electromagnetic form factors:
\begin{equation}
\label{gVPgamma}
\begin{array}{l}
\langle P(p_P)|J_\mu^{\rm EM}|V(p_V,\lambda)\rangle|_{(p_V-p_P)^2=0}=\\[2ex]
\qquad\qquad
=-g_{VP\gamma}\epsilon_{\mu\nu\alpha\beta} p_P^\nu p_V^\alpha\varepsilon_V^\beta(\lambda)\ .
\end{array}
\end{equation}
The decay widths of $P\rightarrow V\gamma$ and $V\rightarrow P\gamma$ are
\begin{equation}
\label{GVPgamma}
\begin{array}{l}
\Gamma(P\rightarrow V\gamma)=
\frac{\alpha}{8}g_{VP\gamma}^2\left(\frac{m_P^2-m_V^2}{m_P}\right)^3\ ,\\[1ex]
\Gamma(V\rightarrow P\gamma)=
\frac{\alpha}{24}g_{VP\gamma}^2\left(\frac{m_V^2-m_P^2}{m_V}\right)^3\ .
\end{array}
\end{equation}

\noindent
Eq.~(\ref{VMD}) allows us to identify the $g_{VP\gamma}$ couplings defined in
(\ref{gVPgamma}) with the form factors $F_{VP\gamma}(0,0)$ listed in (\ref{FVPgamma00}).
The couplings are expressed in terms of the octet and singlet mixing angles $\theta_8$ and $\theta_0$,
the pseudoscalar decay constants $f_8$ and $f_0$, the $\phi$-$\omega$ mixing angle $\theta_V$,
and the corresponding vector decay constants $f_V$.

\noindent
In order to reach some predictions from our two mixing angle analysis
we must first know the values of $\theta_8$ and $\theta_0$
preferred by the experimental data.
We will use as constraints the experimental decay widths of 
$V\rightarrow(\eta,\eta^\prime)\gamma$, and $\eta^\prime\rightarrow V\gamma$, 
with $V=\rho, \omega, \phi$ \cite{Eidelman:2004wy}.
We have performed various fits to this set of
experimental data assuming, or not, the two mixing angle scheme of the $\eta$-$\eta^\prime$ system. 
The results are presented in Table \ref{table1}.

\begin{table}
\centering
\begin{tabular}{cc}
\hline\hline
&\\
Assumptions & Results \\[1ex]\hline
&\\
$\theta_8$ and $\theta_0$ free  & $\theta_8=(-22.3\pm 2.1)^\circ$\\[1ex]
$f_8$ and $f_0$ free                    & $\theta_0=(-1.5\pm 2.4)^\circ$\\[1ex]
$\theta_V=(38.7\pm 0.2)^\circ$  & $f_8=(1.52\pm 0.05) f_\pi$\\[1ex]
                                                         & $f_0=(1.34\pm 0.06) f_\pi$\\[1ex]\hline
&\\
$\theta_8=\theta_0\equiv\theta$ & $\theta=(-15.2\pm 1.4)^\circ$\\[1ex]
$f_8$ and $f_0$ free                    & $f_8=(1.47\pm 0.05) f_\pi$\\[1ex]
$\theta_V=(38.7\pm 0.2)^\circ$  & $f_0=(1.20\pm 0.04) f_\pi$\\[2ex]
\hline\hline
\end{tabular}
\caption{Results for the $\eta$-$\eta^\prime$ mixing angles and decay constants in the octet-singlet basis of the two mixing angle scheme \textit{(up)}
and in the one mixing angle scheme \textit{(down)}.}
\label{table1}
\end{table}

\noindent
As seen from Table \ref{table1}, a significant improvement in the $\chi^2/\textrm{d.o.f.}$~is
achieved when the constrain $\theta_8=\theta_0\equiv\theta$ is relaxed
(the $\chi^2/\textrm{d.o.f.}$~is reduced by more than a factor of 2),
allowing us to show explicitly the improvement of our analysis using the two mixing angle scheme
with respect to the one using the one mixing angle scheme.
It is worth noting that the $\theta_8$ and $\theta_0$ mixing angle values are different at the 
$3\sigma$ level.
Notice also that the experimental data seem to prefer a value for $f_8$ 
higher than the one predicted by $\chi$PT ($f_8=1.28 f_\pi$),
while for the parameters $\theta_8, \theta_0, \theta$ and $f_0$ our values are in agreement with those of
Refs.~\cite{Leutwyler:1997yr,Bramon:1997va}.

\begin{table}
\centering
\begin{tabular}{llll}
\hline\hline\\
$V$ & $P$ & 
$g_{VP\gamma}^{\rm th.}$ (GeV$^{-1}$) & $g_{VP\gamma}^{\rm exp.}$ (GeV$^{-1}$) \\[2ex]
\hline\\
$\rho$ & $\eta$ 
& $(1.34\pm 0.09)$  & $(1.59\pm 0.11)$  \\[1ex]
$\rho$ & $\eta^\prime$ 
& $(1.22\pm 0.09)$  & $(1.35\pm 0.06)$  \\[1ex]
$\omega$ & $\eta$ 
& $(0.51\pm 0.04)$  & $(0.46\pm 0.02)$  \\[1ex]
$\omega$ & $\eta^\prime$
& $(0.55\pm 0.04)$  & $(0.46\pm 0.03)$  \\[1ex]
$\phi$ & $\eta$ 
& $(-0.69\pm 0.05)$ & $(-0.690\pm 0.008)$ \\[1ex]
$\phi$ & $\eta^\prime$
& $(0.70\pm 0.05)$  & $(0.71\pm 0.04)$  \\[2ex]
\hline\hline
\end{tabular}
\caption{Theoretical \textit{(left)} and experimental \textit{(right)} values of the $VP\gamma$ form factors
in the octet-singlet $\eta$-$\eta^\prime$ mixing angle scheme.}
\label{table2}
\end{table}

\noindent
We also include a numerical prediction for each coupling that should be compared with the 
experimental values extracted from (\ref{GVPgamma}) and Ref.~\cite{Eidelman:2004wy}.
In the numerical analysis we have taken into account a value for the vector mixing angle of
$\theta_V=(38.7\pm 0.2)^\circ$ \cite{Dolinsky:vq}. 
Our predictions are obtained from the results in Table \ref{table1}.
The error quoted in Table \ref{table2} does not reflect the full theoretical uncertainty,
but namely propagates the errors from the mixing parameters ($f_8, f_0, \theta_8$ and $\theta_0$)
and the vector decay constants $f_V$.
The agreement between our theoretical predictions and the experimental values is quite
remarkable with exceptions in the $\omega\eta\gamma$ and $\omega\eta^\prime\gamma$ cases.
However, these two couplings merit some explanation.
On one case, the experimental value for $g_{\omega\eta\gamma}$ has changed from 
$(0.53\pm 0.05)$ GeV$^{-1}$ of the PDG'02 \cite{Hagiwara:2002fs} to the current
$(0.46\pm 0.02)$ GeV$^{-1}$ due to the exclusion of the measurement based on 
$e^+e^-\to\eta\gamma$ by Dolinsky \textit{et.~al.}~\cite{Dolinsky:vq}.
On the other case, the $g_{\omega\eta^\prime\gamma}$ coupling is rather sensitive to the
$\phi$-$\omega$ mixing angle;
for instance setting $\theta_V$ to the ideal mixing value of $35.3^\circ$
reduces the coupling by a 10\%.
As seen from Table \ref{table2}, the predictions for the decays $\phi\to (\eta,\eta^\prime)\gamma$,
which are the best measured and are largely independent of $\theta_V$,
are in good agreement with data.
For completeness, we have to mention that fixing $\theta_8=\theta_0\equiv\theta$
the coupling $g_{\phi\eta^\prime\gamma}$ is fitted to $(0.98\pm 0.05)$ GeV$^{-1}$
which is in clear contradiction with data.

\noindent
Table \ref{table2} constitutes one of the main results of our work.
Our analysis shows that the assumption of saturating the form factors
$F_{VP\gamma}$ by lowest-lying resonances is satisfactory (a conclusion 
already reached in Ref.~\cite{Ball:1995zv}),
and that the $\eta$-$\eta^\prime$ system described in the two mixing angle scheme
(octet-singlet basis) fits the data much better than the one mixing angle scheme does.

\begin{table}
\centering
\begin{tabular}{cc}
\hline\hline
&\\
Assumptions & Results\\[1ex]\hline
&\\
$\phi_q$ and $\phi_s$ free   & $\phi_q=(42.0\pm 2.3)^\circ$\\[1ex]
$f_q$ and $f_s$ free              & $\phi_s=(42.3\pm 2.0)^\circ$\\[1ex]
$\phi_V=(3.4\pm 0.2)^\circ$ & $f_q=(1.13\pm 0.04) f_\pi$\\[1ex]
                                                  & $f_s=(1.68\pm 0.07) f_\pi$\\[1ex]\hline
&\\
$\phi_q=\phi_s\equiv\phi$     & $\phi=(42.2\pm 1.5)^\circ$\\[1ex]
$f_q$ and $f_s$ free              & $f_q=(1.13\pm 0.04) f_\pi$\\[1ex]
$\phi_V=(3.4\pm 0.2)^\circ$ & $f_s=(1.68\pm 0.07) f_\pi$\\[2ex]
\hline\hline
\end{tabular}
\caption{Results for the $\eta$-$\eta^\prime$ mixing angles and decay constants
in the quark-flavour basis.
The conventions are the same as in Table \protect\ref{table1}.}
\label{table3}
\end{table}

\noindent
Up to now, we have shown in the octet-singlet basis the need for a two mixing angle scheme
in order to describe experimental data in a better way.
In the following, we proceed to perform the same kind of analysis but in the quark-flavour basis.
In Table \ref{table3} we present the results of the fits taking into account
the experimental data available
---namely $V\to P\gamma$ and $P\to V\gamma$---
and the corresponding theoretical expressions found in the Appendix of Ref.~\cite{Escribano:2005qq}.
As seen from Table \ref{table3}, there is no significant difference at the 
\emph{current experimental} accuracy between the
$\chi^2/\textrm{d.o.f.}$ of the fits when data are described in terms of 
two mixing angles (in the quark-flavour basis) or if  $\phi_q=\phi_s\equiv\phi$ is fixed.
There is however no strong reason to impose this equality as a constraint.
Our results are in agreement with those of Ref.~\cite{Feldmann:1999uf}
except for the large value of $f_s$  which is forced by the $\phi\to\eta^\prime\gamma$ decay
not included in the previous analysis.

\section{CONCLUSIONS}
\noindent
We have performed a phenomenological analysis on various decay processes
using a two mixing angle scheme for the $\eta$-$\eta^\prime$ system.
The agreement between our theoretical predictions and the experimental values
is remarkable and can be considered as a consistency check of the whole approach.
We have shown that a two mixing angle description in the octet-singlet basis is required in order to achieve good agreement with experimental data.
On the contrary, in the quark-flavour basis and with the present experimental accuracy
a one mixing angle description of the processes is still enough to reach agreement.
This behaviour gives experimental support to the fact that the difference of the
two mixing angles in the octet-singlet basis is a $SU(3)$-breaking effect 
while in the quark-flavour basis is a OZI-rule violating effect which appears to be smaller.


\begin{thebibliography}{99}
\bibitem{Ball:1995zv}
P.~Ball, J.~M.~Frere and M.~Tytgat,
Phys.\ Lett.\ B {\bf 365}, 367 (1996)
[arXiv:hep-ph/9508359].

\bibitem{Kaiser:2000gs}
R.~Kaiser and H.~Leutwyler,
Eur.\ Phys.\ J.\ C {\bf 17} (2000) 623
[arXiv:hep-ph/0007101].

\bibitem{Leutwyler:1997yr}
H.~Leutwyler,
Nucl.\ Phys.\ Proc.\ Suppl.\  {\bf 64} (1998) 223
[arXiv:hep-ph/9709408].

\bibitem{Kaiser:1998ds}
R.~Kaiser and H.~Leutwyler,
arXiv:hep-ph/9806336.

\bibitem{Feldmann:1999uf}
T.~Feldmann,
Int.\ J.\ Mod.\ Phys.\ A {\bf 15} (2000) 159
[arXiv:hep-ph/9907491].

\bibitem{Eidelman:2004wy}
S.~Eidelman {\it et al.}  [Particle Data Group Collaboration],
Phys.\ Lett.\ B {\bf 592} (2004) 1.

\bibitem{Bramon:1997va}
A.~Bramon, R.~Escribano and M.~D.~Scadron,
Eur.\ Phys.\ J.\ C {\bf 7}, 271 (1999)
[arXiv:hep-ph/9711229];

\bibitem{Dolinsky:vq}
S.~I.~Dolinsky {\it et al.},
Phys.\ Rept.\  {\bf 202} (1991) 99.

\bibitem{Hagiwara:2002fs}
K.~Hagiwara {\it et al.}  [Particle Data Group Collaboration],
Phys.\ Rev.\ D {\bf 66} (2002) 010001.

\bibitem{Escribano:2005qq}
  R.~Escribano and J.~M.~Frere,
  JHEP {\bf 0506} (2005) 029
  [arXiv:hep-ph/0501072].

\end{thebibliography}
\end{document}